\documentclass{ecai} 

\usepackage{latexsym}
\usepackage{amssymb}
\usepackage{amsmath}
\usepackage{amsthm}
\usepackage{booktabs}
\usepackage{enumitem}
\usepackage{graphicx}
\usepackage{color}
\usepackage{microtype}
\usepackage{hyperref}

\usepackage[utf8]{inputenc}
\usepackage[T1]{fontenc}

\usepackage{amsmath}
\usepackage{amsthm}
\usepackage{amssymb}
\usepackage{amsfonts}

\usepackage{booktabs}
\usepackage{tabularx}

\usepackage{placeins}
\usepackage{enumitem}

\usepackage{wrapfig} 

\usepackage{algorithm}
\usepackage{algorithmic}

\newcommand{\BibTeX}{B\kern-.05em{\sc i\kern-.025em b}\kern-.08em\TeX}

\usepackage{cleveref}
\crefname{figure}{figure}{figures}

\hypersetup{
		pdfencoding=auto, 
		psdextra,
		colorlinks=true,
		citecolor=green!40!black,
		linkcolor=red!50!black,
		urlcolor=blue!80!black
	}
\usepackage{subcaption}

\let\oldemph\emph
\renewcommand{\emph}[1]{\textcolor{blue!60!black}{\oldemph{#1}}}

\usepackage{fancyhdr}
\usepackage{mathtools}
\usepackage{pifont}
\usepackage{subcaption}
\usepackage{booktabs}
\usepackage{multirow}
\usepackage[export]{adjustbox}
\usepackage{xargs}
\usepackage{xcolor}

\usepackage{booktabs}
\usepackage{nicefrac}
\usepackage{chngpage}
\usepackage{todonotes}
\usepackage{cleveref}
\usepackage{placeins}

\usepackage{tikz}
\usetikzlibrary{cd}

\newtheorem{theorem}{Theorem}

\newtheorem{corollary}[theorem]{Corollary}

\newtheorem{definition}{Definition}

\theoremstyle{remark}
\newtheorem{remark}{Remark}[section]
\theoremstyle{plain}

\newtheorem{method}{Rule}

\numberwithin{equation}{section}

\DeclareMathOperator{\relint}{ri}
\DeclareMathOperator{\relbd}{rbd}

\newcommand{\normphi}{{{\mathrm{norm}\hbox{-}\phi}}}

\usepackage{todonotes}

\makeatletter 
\def\WFfill{\par 
    \ifx\parshape\WF@fudgeparshape 
    \nobreak 
    \ifnum\c@WF@wrappedlines>\@ne 
    \advance\c@WF@wrappedlines\m@ne 
    \vskip\c@WF@wrappedlines\baselineskip 
    \global\c@WF@wrappedlines\z@ 
    \fi 
    \allowbreak 
    \WF@finale 
    \fi 
} 
\makeatother

\newcolumntype{R}{>{\raggedleft\arraybackslash}X}
\newcolumntype{Z}{>{\centering\arraybackslash}X}
\newcolumntype{W}{>{\centering\arraybackslash}m{.75in}}

\begin{document}


\begin{frontmatter}


\paperid{7030} 

 
\title{Spoiler Susceptibility in Party Elections}


\author[A]{\fnms{Daria}~\snm{Boratyn}\orcid{0000-0003-3299-7071}}
\author[A]{\fnms{Wojciech}~\snm{Słomczyński}\orcid{0000-0003-2388-8930}}
\author[A]{\fnms{Dariusz}~\snm{Stolicki}\orcid{0000-0002-8295-0848}\thanks{Corresponding Author. Email: dariusz.stolicki@uj.edu.pl}}
\author[B,C]{\fnms{Stanis\l{}aw}~\snm{Szufa}\orcid{0000-0001-6301-6227}} 

\address[A]{Jagiellonian University, Center for Quantitative Political Science, Kraków, Poland}
\address[B]{CNRS, LAMSADE, Universit\'{e} Paris Dauphine -- PSL, Paris, France}
\address[C]{AGH University, Krak\'ow, Poland}

\begin{abstract}
An electoral spoiler is usually defined as a losing candidate whose removal would affect the outcome by changing the winner. So far, spoiler effects have been analyzed primarily for single-winner electoral systems. We consider this subject in the context of party elections, 
where there is no longer a sharp distinction between winners and losers. Hence, we propose a more general definition, under which a party is a spoiler if their elimination causes any other party’s share in the outcome to decrease.

We characterize spoiler-proof electoral allocation rules for zero-sum voting methods. In particular, we prove that for seats-votes functions only identity is spoiler-proof. 
However, even if spoilers are unavoidable under common electoral rules, their expected impact can vary depending on the rule.
Hence, we introduce a measure of spoilership,
which allows us to experimentally compare a number of multiwinner social choice rules according to their spoiler susceptibility. 
Since the probabilistic models used in COMSOC have been developed for nonparty elections, we extend them to generate multidistrict party elections.
\end{abstract}

\end{frontmatter}


\section{Introduction}\label{sec:intro}

In the context of single-winner elections, a \emph{spoiler} is usually defined as a losing candidate whose removal would affect the outcome by changing the winner \citep{Borgers10,Kaminski18}.
In this paper, we extend the definition of a spoiler to \emph{party elections}, investigate spoiler-proofness of multiwinner electoral rules, and analyze their spoiler susceptibility.

We primarily seek to investigate spoiler effects in political elections to multi-member representative bodies. They are distinguished from other multiwinner elections by their character as party elections. By party election we mean such an election where the allocation of some payoff (e.g., \emph{parliamentary seats}) among \emph{parties} (rather than the identity of the winners) is the main outcome.\footnotemark \hspace{0.01em} Thus, spoiler effects should be considered in terms of party seat allocations.

\footnotetext{This definition differs from those put forward by earlier authors addressing party elections in COMSOC, such as \citet{Botan21}, for whom their defining characteristic was a block voting pattern. It instead generalizes the class of apportionment methods, see, e.g., \citet{BredereckEtAl20} and \citet{BrillEtAl22}.}

\subsection{Contribution}


\textbf{First.} We generalize the definition of a spoiler to party elections: A party is deemed a spoiler if there exists another party whose payoff decreases if the former is eliminated.

\textbf{Second.} We prove a general impossibility theorem regarding the existence of spoiler-proof electoral rules that can be described by seats-votes functions distinct from identity.

\textbf{Third.} We characterize spoiler-proof electoral rules for zero-sum voting methods like ordinal or cumulative voting.

\textbf{Fourth.} We compare common apportionment methods, viz., FPTP, Jefferson--D'Hondt, Webster--Sainte-Laguë, and Largest Remainders, according to their expected spoiler susceptibility. We conclude that spoiler susceptibility is negatively correlated with proportionality and with the use of Borda scores instead of first preferences.

\textbf{Fifth.} Since party elections under ordinal voting have not been previously considered in COMSOC, we propose probabilistic models of party elections extending several common statistical cultures.

\subsection{Related Work}

While spoiler effects have long been a familiar subject in the field of voting theory, there have been few attempts to formally define spoilers or to measure the immunity of electoral systems to spoilers \citep[see, e.g.,][]{McCuneWilson23,McCuneWilson24,HollidayPacuit23}. However, spoiler effects have been tangentially considered in classical social choice theory in the context of stronger postulates such as independence of irrelevant alternatives \citep{Arrow50,Ray73,BordesTideman91} or candidate stability \citep{DuttaEtAl01,EhlersWeymark03,EraslanMcLennan04,Rodriguez-Alvarez06}
or distinct though related postulates such as the independence of clones \citep{Tideman87}\footnotemark.

\footnotetext{A clone is a candidate indistinguishable from some candidate $c$ in the original set of candidates (in the sense that for every voter if that voter prefers some candidate $x$ over $c$, they also prefer $x$ over the clone, and vice versa), and preference between $c$ and the clone is random. The motivation for studying clones is that a clone is a natural spoiler for $c$ (and we have a simple model of how preferences change with the addition or deletion of a clone). However, clones are a very specific class of spoilers, making spoiler and clone effects quite distinct research subjects.}

In computational social choice, spoilers have been addressed from the point of view of electoral control problems \citep{MeirEtAl08,LiuEtAl09,ChevaleyreEtAl10,FaliszewskiRothe16,NevelingEtAl20,ErdelyiEtAl21}, in particular of the problems of \textsc{Constructive-Control-by-Adding-Candidates} (CCAC) and \textsc{Constructive-Control-by-Deleting-Candidates} (CCDC), and their destructive control counterparts (DCAC and DCDC). In constructive control problems we seek to alter the election outcome by making a specific candidate a unique winner, while in destructive control problems we seek to make that candidate a non-winner.
In CAC problems that objective is to be achieved by adding spoiler candidates, while in CDC problems we restrict the set of candidates by deleting some.
While closely related to our subject, those problems are distinct, as they treat the election outcome in binary terms 
which are inapposite in party elections. Moreover, research on control problems considers vulnerability to spoilers in terms of computational complexity of solving the control problems, while we focus on the expected impact of a spoiler.

\citet{Kaminski18} has been the first to investigate the concept of spoilers in party elections. He distinguished several ways in which a potential spoiler can affect the seat payoff. Thus, apart from classical spoilers (who turn a majority winner into a majority loser, while making another player a majority winner), he distinguishes kingmaker spoilers (who turn a majority loser into a majority winner), kingslayer spoilers (who turn a majority winner into a majority loser), and valuegobblers (who affect the seat payoff of any one player by an amount greater than their own payoff). However, Kaminski makes no formal assumptions about vote redistribution (and without such assumptions, any redistribution result is attainable -- so we can only speak of a spoiler in a given case, and not of spoiler susceptibility or spoiler-proofness in general), nor does he go beyond empirical examples to consider spoiler-proofness as a more general property of voting rules (in fact, he only deals with one -- Jefferson--D'Hondt). Additionally, his analysis is focused solely on single-choice voting methods.

\section{Introductory Example}\label{sec:example}

While no formal definition of spoilers in party elections exists, psephologists have nevertheless regarded some parties as spoilers. For example, \emph{Reform UK} was widely regarded as a spoiler in the 2024 United Kingdom general election. We shall examine this example more closely to illustrate the intuition underlying our proposed definition of a spoiler.


In the United Kingdom, first-past-the-post (plurality) voting is used to allocate a total of $650$ seats in single-member districts. In 2024, there have been five major national contenders, see Table \ref{tbl:ukParties}, columns 1 to 3 ($37$ seats went to regional and minor parties, which are of no interest to us). Let us consider what would happen if \emph{Reform UK} had not participated in the election. Its votes would likely have been redistributed among other parties. Assume for the sake of argument that two-thirds of them would go to the Conservatives, as the ideologically closest major party, while the remainder would stay at home.\footnotemark

\footnotetext{We only use this assumption in the present example for illustrative purposes. Throughout the remainder of this article, we admit arbitrary vote redistribution functions satisfying Def. \ref{def:restrict}.}

We would have expected that since every party gains votes when \emph{Reform} is removed, their seat shares should not decrease. However, an inspection of the results reveals that two parties lose seats (including Labour, which no longer commands a majority), and $122$ seats ($18.77\%$) change hands, even though \emph{Reform} held only $5$ (this happens since, as Reform votes get redistributed, losing Conservative candidates can receive enough to overtake winning Labour candidates in their district, leading to Labour losing seats without losing any votes). We argue that this effect, i.e., a party's removal causing another to lose seat shares, is the essence of spoilership. 

\begin{table}[h]
\newcolumntype{Z}{>{\centering\arraybackslash}X}
\begin{tabularx}{\columnwidth}{lZZZZ} \toprule
\multirow{2}{*}{name} & \multicolumn{2}{>{\hsize=\dimexpr2\hsize+2\tabcolsep+\arrayrulewidth\relax}Z}{actual} & \multicolumn{2}{>{\hsize=\dimexpr2\hsize+2\tabcolsep+\arrayrulewidth\relax}Z}{no \emph{Reform UK}} \\
 & votes & seats & votes & seats \\ \midrule
Labour & $.337$ & $.632$ & $.354$ & $.478$ \\
Conservatives & $.237$ & $.186$ & $.349$ & $.374$ \\
Reform UK & $.143$ & $.008$ & $.000$ & $.000$ \\
Lib. Dems. & $.122$ & $.111$ & $.128$ & $.089$ \\
Greens & $.067$ & $.006$ & $.070$ & $.006$ \\
\bottomrule
\end{tabularx}
\caption{United Kingdom general election, 2024.}
\label{tbl:ukParties}
\end{table}

Of course, plurality voting in single-winner elections is widely reputed to be highly spoiler-susceptible \citep[Sec.~2]{McCuneWilson23}, so perhaps a switch to another voting system would eliminate spoilers. However, examples of spoilers can also be found in countries using proportional representation systems. For instance, a small left-wing party \emph{Razem} was a spoiler in the Polish general election of 2015, held under the Jefferson--D'Hondt rule with a $5\%$ statutory threshold for parties and $8\%$ statutory thresholds for coalitions. If \emph{Razem} were removed from that election, and its votes redistributed in proportion to ideological proximity \citep{BakkerEtAl21}, four parties would lose seats, the election winner, PiS, would no longer command a majority, and $9.5\%$ of seats would change hands, even though \emph{Razem} held none (see Table \ref{tbl:plParties} for details).

\begin{table}[h]
\newcolumntype{Z}{>{\centering\arraybackslash}X}
\begin{tabularx}{\columnwidth}{lZZZZZZ} \toprule
\multirow{2}{*}{name} & \multicolumn{2}{>{\hsize=\dimexpr2\hsize+2\tabcolsep+\arrayrulewidth\relax}Z}{actual} & \multicolumn{2}{>{\hsize=\dimexpr2\hsize+2\tabcolsep+\arrayrulewidth\relax}Z}{no \emph{Razem}} \\ 
 & votes & seats & votes & seats \\ \midrule 
PiS & $.376$ & $.511$ & $.380$ & $.454$ \\ 
PO & $.241$ & $.300$ & $.246$ & $.278$ \\ 
Kukiz & $.088$ & $.091$ & $.092$ & $.076$ \\ 
Nowoczesna & $.076$ & $.061$ & $.081$ & $.062$ \\ 
Lewica (*) & $.076$ & $.000$ & $.085$ & $.067$ \\ 
PSL & $.051$ & $.035$ & $.056$ & $.029$ \\ 
Korwin & $.048$ & $.000$ & $.051$ & $.022$ \\ 
Razem & $.036$ & $.000$ & $.000$ & $.000$ \\ 
\bottomrule
\end{tabularx}
\caption{Polish general election, 2015. Asterisk (*) marks coalitions.}
\label{tbl:plParties}
\end{table}


\section{Preliminaries}

We begin with the definition of two fundamental concepts underlying our definition of an election. First, let $P$ be a set of candidates (\emph{parties}). For convenience, we assume $P$ is a finite subset of $\mathbb{N}$, i.e., we identify parties with their indices. Second, let $\mathcal{D}_{P}$ denote the set of all admissible votes. How exactly that set is defined depends on the specific voting method used in the election (see Table \ref{tbl:votingMethods} for examples).

\begin{table}[H]
\newcolumntype{Z}{>{\centering\arraybackslash}X}
\begin{tabularx}{\columnwidth}{l|X}
    method & $\mathcal{D}_{P}$ \\ \hline
    single-choice & $\{\mathbf{x} \in \{0,1\}^{P}: \lVert\mathbf{x}\rVert_1 = 1\}$ (isomorphic to $P$) \\
    ordinal & linear orders on $P$ \\
    approval & $\{0, 1\}^{P}$ \\
    cumulative & $\Delta_P$ \\
    cardinal & $[0, 1]^P$
\end{tabularx}
\caption{Domains of admissible votes for common voting methods.}
\label{tbl:votingMethods}
\end{table}

\setlength{\belowcaptionskip}{2pt}

For the sake of simplicity, we restrict ourselves to voting methods where votes can be represented either by total preorders (\emph{ranked voting methods}) or by vectors of real numbers (\emph{cardinal voting methods}). This class includes virtually all of the voting methods that are of interest in computational social choice (including single-choice, ordinal, approval, scoring, cumulative, and knapsack voting). However, this restriction is just to simplify formalism. The concepts themselves can be applied for any voting method whatsoever.

\begin{definition}[Election]
    An \emph{election} ($P$, $\mathbf{w}$) is a pair consisting of a set of parties, $P$, and a \emph{profile} (i.e., collection of votes), $\mathbf{w}$. We denote the set of all profiles of votes from $\mathcal{D}_{P}$ by $\mathcal{V}_{P}$.
\end{definition}

As in the case of multi-winner elections, what distinguishes party elections among other kinds of elections is a special type of electoral rules. \emph{Party elections} are those where the electoral rule is an \emph{allocation rule}, mapping parties to their payoffs. In the canonical examples, those payoffs correspond to legislative seat shares, but the model itself is much more general: They can as well represent budget shares in a participatory budgeting system, voting power indices, etc. It is convenient to normalize those payoffs so that they add up to $1$, ensuring that the payoff vector is a point on some $m$-dimensional unit (probability) simplex, where $m$ is the number of parties, and coordinates correspond to parties:

\begin{definition}[Allocation Rule]\label{def:allocRule}
    An \emph{allocation rule} is a function that maps a profile $\mathbf{w} \in \mathcal{V}_{P}$ to a payoff vector $\mathbf{s} \in \Delta_{P}$ for every $P \subset \mathbb{N}$.
\end{definition}

It is technically convenient to define the space of payoff vectors more generally, as an infinitely-dimensional unit simplex whose coordinates correspond to all possible parties. This is not to say that we require the number of parties to be infinite: We just avoid the need to fix the maximum number of parties when defining our electoral rules, while leaving open the option of using virtual parties as a proof technique (as in Theorem \ref{thm:nogo}). Accordingly, we denote our universal payoff space as $\Delta := \{ \mathbf{x} \in \mathbb{R}_{+}^{\mathbb{N}} : \sum_{i \in \mathbb{N}} x_i = 1\}$. For any non-empty set of natural numbers, $S \subset \mathbb{N}$, we denote the face of $\Delta$ for which only the coordinates in $S$ are non-zero by $\Delta_{S}:=\{ \mathbf{x} \in \Delta : \sum_{i \in S} x_i = 1 \}$. Finally, if $S$ is given, for every $i \in S$ we denote $\Delta_{S \setminus \{i\}}$ by $\Delta_{-i}$.

Finally, let us introduce some more technical mathematical notation that will be used throughout the paper:
\begin{description}
    \item[$\mathcal{P}$] For any set $X$, we denote its power set by $\mathcal{P}(X)$.
    \item[$\restriction$] For any relation $r$ on a set $X$ (function, order, etc.), we denote its \textbf{restriction} to some $Y \subset X$ by $r \restriction_{Z}$.
    \item[$\lVert\cdot\rVert_{1}$] For any vector $\mathbf{x} \in \mathbb{R}^{n}$, where $n = 1, \dots, \omega$, we denote its $L_1$ (absolute value) norm by $\lVert\mathbf{x}\rVert_{1}$.
\end{description}
In general, we use capital letters to denote sets and random variables, and boldface font for vectors, votes, and profiles.

\section{Definition of a Spoiler}\label{sec:def}


The fact that a party's election outcome is defined in terms of a payoff in $[0, 1]$ complicates the traditional single-winner definition of a spoiler, as there is no longer a sharp distinction between losers and winners. Clearly, a loser who affects the payoff distribution is the closest analogue of a spoiler, but a question remains of how to distinguish one. Defining a loser as a party with the payoff zero seems too restrictive, as under many real-life allocation rules even very small parties can get some non-zero seat share. But if we admit that a party with a positive payoff can potentially be a spoiler, the traditional definition becomes unworkable: If the payoffs are to add up to $1$, eliminating any party with a non-zero payoff entails a change of the payoff of at least one other party.

However, we would expect this change not to go beyond the reallocation of the payoff of the party being eliminated. In other words, no payoffs originally allocated to other parties should be reallocated. However, as the payoffs lie on a unit simplex, if the total measure of the reallocated payoff exceeds the share of the party being eliminated, then necessarily some other party's payoff also decreases. This brings to our definition of a spoiler in party elections, stated at first informally: {\em A party is a spoiler if and only if its elimination causes a change in party payoffs other than through the reallocation of its own payoff (i.e., if it causes a decrease in some other party's payoff)}.

Formally, a \emph{payoff reallocation} should be a \emph{monotonic projection}, i.e., a projection of a point in a unit simplex onto a face of that simplex such that every coordinate either decreases to zero or not at all:

\begin{definition}[Monotonic Projection]\label{def:monoProj}
    For any $P \in \mathcal{P}(\mathbb{N})$ and any $\emptyset \neq K \subseteq P$, a function $\pi: \Delta_{P} \rightarrow \Delta_{K}$ is a \emph{monotonic projection} if $\pi^2 = \pi$ and for all $\mathbf{x} \in \Delta_{P}$ and $j \in K$ we have that $\pi_j(\mathbf{x}) \ge x_j$.
\end{definition}

\noindent For $i \in P$, the set of all monotonic projections $\Delta_{P} \rightarrow \Delta_{P \setminus \{i\}}$ is denoted by $\Pi_{P,-i}$.\smallskip

We also need to define the \emph{vote redistribution function}, a mathematical model of the operation of eliminating a party. Our basic assumption here is that it satisfies the \emph{Individual Independence of Irrelevant Alternatives} axiom \citep{RadnerMarschak54}: How each individual voter ranks or scores a set of parties should not depend on what other alternatives exist or how they are ranked or scored.

\begin{definition}[Individual Independence of Irrelevant Alternatives]\label{def:iiia}
    For any $P \in \mathcal{P}(\mathbb{N})$ and any $\emptyset \neq K \subseteq P$, a function (\emph{vote transform}) $\psi: \mathcal{D}_{P} \rightarrow \mathcal{D}_{K}$ satisfies \emph{Individual Independence of Irrelevant Alternatives} if:
    \begin{itemize}
        \item for every $\mathbf{v} \in \mathcal{D}_{P}$ if the restriction of $\mathbf{v}$ to $K$, $\mathbf{v} {\restriction}_{K}$, is an element of $\mathcal{D}_{K}$, then $\psi(\mathbf{v}) = \mathbf{v} {\restriction}_{K}$,
        \item for ranked voting methods, $i \prec_{\mathbf{v}} j$ implies $i \prec_{\psi(\mathbf{v})} j$ for every $\mathbf{v} \in \mathcal{D}_{P}$ and every $i, j \in K$,
        \item for cardinal voting methods, $\psi(\mathbf{v})_i \ge \mathbf{v}_i$ for every $\mathbf{v} \in \mathcal{D}_{P}$ and every $i \in K$.
    \end{itemize}
\end{definition}

Clearly, it would be unrealistic to model party removal by a function that deterministically transforms every vote in the same manner. Since elements of a collection are indiscernible, we employ a stochastic approach. Let $\mathcal{T}$ be some probability distribution on the set of all vote transforms. A \emph{stochastic vote transform} is an elementwise function that maps the $i$-th vote\footnotemark in a profile $\mathbf{w}$ to its image under a realization of a random vote transform $\varphi_i$, where $\varphi_1, \dots, \varphi_{|\mathbf{w}|} \sim \mathcal{T}$ are i.i.d. It satisfies individual IIA if the set of all vote transforms that do not satisfy individual IIA is a null set under $\mathcal{T}$.

\footnotetext{Even if the profile is infinite, we can index the votes with ordinals per the well-ordering theorem.}

For example, consider single-choice voting. Let $P = \{1, 2, 3\}$. When removing the third party, there are two vote transforms (up to isomorphism): one maps $3$ to $1$, and the other maps $3$ to $2$. Realistically, a redistribution of votes would be a mix of the two. Assume that a vote for $3$ is transformed to a vote for $1$ with probability $p$, and to a vote for $2$ with probability $1-p$. We redistribute each vote randomly according to those probabilities (this is the stochastic element).

\begin{definition}[Vote Redistribution Function]\label{def:restrict}
    For any $P \in \mathcal{P}(\mathbb{N})$ and any $\emptyset \neq K \subseteq P$, a vote redistribution function $\rho: \mathcal{V}_{P} \rightarrow \mathcal{V}_{K}$ is any stochastic vote transform that satisfies \emph{Individual Independence of Irrelevant Alternatives}.
\end{definition}

\noindent We denote the set of all vote redistribution functions on $P$ to $K$ by $R_{P,K}$. For any $i \in P \subset \mathbb{N}$, let $R_{P,-i} := R_{P, P \setminus \{i\}}$.\smallskip

Under some voting methods, such as approval or ordinal voting, for every $P$ and $K$ the restriction of any vote from $P$ to $K$ is a valid vote: We just omit all approvals for a given party or delete from a preference ranking, and the vote remains valid. In such a case, such a restriction defines a unique redistribution function. But under cumulative or single-choice voting, for instance, we need the eliminated party's votes to be redistributed, and there are multiple ways of doing so, corresponding to multiple vote redistribution functions. For instance, in our introductory example, we have redistributed votes in proportion to inverse distance, but we could also have done it in proportion to party size, equally, or in infinitely many other ways, each of which is a different vote redistribution function.

We now have introduced all the concepts necessary for a definition of an electoral spoiler:

\begin{definition}[Electoral Spoiler]\label{def:spoiler}
    Let $\emptyset \neq P \subset \mathbb{N}$ and let $\mathbf{w} \in \mathcal{V}_{P}$ be a profile. Then the $i$-th party, $i \in P$, is an \emph{electoral spoiler} under an allocation rule $f$ and a vote redistribution function $\rho \in R_{P,-i}$ if and only if there exists no monotonic projection $\pi \in \Pi_{P,-i}$ such that $f(\rho(\mathbf{w})) = \pi(f(\mathbf{w}))$.
\end{definition}

Let us briefly consider the intuition behind that definition. We have posited that a party is a spoiler if and only if its elimination causes a change in payoffs beyond the reallocation of its own payoff. We can now reformulate that condition in more formal terms: A party is a spoiler unless the payoff reallocation incident to its elimination is a monotonic projection. In other words, there should exist some monotonic projection $\pi$ that forms the following commutative diagram with the allocation rule and the vote redistribution function:

\begin{equation*}
\begin{tikzcd}[row sep=large,column sep=large,every label/.append style = {font = \normalsize}]
\mathbf{w} \in \mathcal{V}_{P}
    \arrow[r, "f"]
    \arrow[d, "\rho"] &
f(\mathbf{w}) \in \Delta_{P}
    \arrow[d, "\pi"] \\
\rho(\mathbf{w}) \in \mathcal{V}(P {\setminus} \{i\})
    \arrow[r, "f"] &
f(\rho(\mathbf{w})) \in \Delta_{P {\setminus} \{i\}}
\end{tikzcd}
\end{equation*}

\noindent A spoiler is a party for which no such projection exists. In such a case, as the payoff reallocation is not a monotonic projection, it involves a payoff decrease for some party:

\begin{corollary}
    If the $i$-th party is a spoiler under $\rho \in R_{P,-i}$, then there exists a party $j \neq i$ such that $f_j(\rho(\mathbf{w})) < f_j(\mathbf{w})$, i.e., $j$'s payoff decreases when $i$ is eliminated.
\end{corollary}

\noindent We can now proceed to define spoiler-proofness:

\begin{definition}[Spoiler-Proofness]\label{def:spoilProof}
    An allocation rule $f$ is \emph{spoiler-proof} if for every non-empty $P \subset \mathbb{N}$, profile $\mathbf{w} \in \mathcal{V}_{P}$, and vote redistribution function $\rho \in R_{P,-i}$ no party $i \in P$ is a spoiler for such $f$, $\mathbf{w}$, and $\rho$.
\end{definition}

\begin{figure}[h]
    \centering
    \resizebox{\columnwidth}{!}{
    \begin{tikzpicture}[scale=5]
    \draw (0,0) -- (0.5,0.866025) -- (1,0) -- cycle;
    \filldraw[black] (0.361111,0.336788) circle (0.2pt) node[anchor=north]{$f(\mathbf{w})$};
    \filldraw[black] (0.805556,0.336788) circle (0.2pt) node[anchor=west]{$f(\mathbf{w}) + f_i(\mathbf{w}) (w_1 - w_i)$};
    \filldraw[black] (0.583333,0.721688) circle (0.2pt) node[anchor=west]{$f(\mathbf{w}) + f_i(\mathbf{w}) (w_2 - w_i)$};
    \filldraw[black] (0.875,0.216506) circle (0.2pt) node[anchor=west]{$f(\rho(\mathbf{w}))$};
    \filldraw[black] (0,0) circle (0pt) node[anchor=north east]{$(0, 0, 1)$};
    \filldraw[black] (1,0) circle (0pt) node[anchor=north west]{$(1, 0, 0)$};
    \filldraw[black] (0.5,0.866025) circle (0pt) node[anchor=south]{$(0, 1, 0)$};
    \draw[gray,ultra thick] (0.805556,0.336788) -- (0.583333,0.721688);
    \draw[black] (0.694444,0.529238) node[anchor=west]{$\{\pi(f(\mathbf{w})): \pi \in \Pi_{P,-i}\}$};
    \draw[very thick] (0.805556,0.336788) -- (0.875,0.216506);
    \draw[black] (0.847222, 0.254619) node[anchor=east]{$\lambda_i$};
    \draw[thin,dashed] (0.361111,0.336788) -- (0.805556,0.336788);
    \draw[thin,dashed] (0.361111,0.336788) -- (0.583333,0.721688);
    \end{tikzpicture}
    }
    \caption{Geometric interpretation of Definitions \ref{def:spoiler} and \ref{def:spoilership}.}
    \label{fig:elecImpact}
\end{figure}

However, just as party elections are characterized by the non-Booleanness of their outcome, the spoiler status in such elections is similarly non-binary, as the excess payoff reallocation caused by a spoiler's elimination can vary in size. Our framework enables us to quantify that variation by yielding a natural measure thereof:

\begin{definition}[Excess Electoral Impact]\label{def:spoilership}
    Let a set of parties $\emptyset \neq P \subset \mathbb{N}$ and a profile $\mathbf{w} \in \mathcal{V}_{P}$ be given. By $\lambda_{i, \rho}^f$, where $\rho \in R_{-i}$, we denote the \emph{excess electoral impact} of the $i$-th party under an allocation rule $f$ and vote redistribution function $\rho$, i.e., the minimal $L_1$ distance between $f(\rho(\mathbf{w}))$ and the \emph{reallocation region} $\{\pi(f(\mathbf{w})): \pi \in \Pi_{P,-i}\}$.
\end{definition}

\begin{remark}
    Note that, for a given allocation rule $f$ and vote redistribution function $\rho$, the excess electoral impact $\lambda_{i,\rho}^f > 0$ if and only if the party $i$ is a spoiler under $f$ and $\rho$.
\end{remark}

Thus, in our introductory examples from Sec. \ref{sec:example}, excess electoral impact of \emph{Reform UK} in the 2024 UK election equals $.364$, while excess electoral impact of \emph{Razem} in the 2015 Polish election is $.190$.

The above definitions are illustrated by \Cref{fig:elecImpact}). The simplex is the space of all payoff allocations under a three-party election. The point described as $f(\mathbf{w})$ is the allocation under the original profile of votes. The subset of the face between $(0,1,0)$ and $(1,0,0)$ marked by bolder gray line is the reallocation region under the removal of the third party. The point described as $f(\rho(\mathbf{w}))$ is the allocation under the removal of the third party and redistribution of its votes described by function $\rho$. It falls outside the reallocation region, so the third party is a spoiler, and $\lambda_i$ is the measure of its excess electoral impact.

\section{Seats-Votes Elections}\label{sec:seatsVotes}

In this section, we introduce the class of \emph{seats-votes elections}, i.e., such elections that every profile can be represented as a vector of probabilities (usually representing vote shares). We focus on such systems for several reasons. One is attainability of powerful theoretical results at the cost of relatively lightweight formalism and few technicalities. Another one is the ubiquity of such systems. Quite a few electoral systems used in real-life are seats-votes elections, in the sense that a vector of vote shares contains all the information needed to compute seat allocations \citep{Colomer04}. Moreover, a large majority of real-life electoral systems can be well approximated by seats-votes elections \cite{TaageperaShugart89}. Finally, representing the vote profile in terms of vote shares comes intuitively to most people, as this is how election results or polls are usually reported.

The motivating example behind seats-votes elections are single-choice list-based proportional representation elections, used to elect parliaments in over 60 countries, where each vote consists of a choice of a party. There, a profile can be represented as a vector of vote shares obtained by each party. However, one can also consider a system where voters cast ordinal ballots, but seats are allocated on the basis of the sums of party Borda scores, and so the vector of normalized scores is the profile. Finally, a profile may have no interpretation in terms of individual votes, e.g., a weighted voting game with an allocation rule that maps a vector of voting weights to a vector of voting powers.

Formally, an election is a \emph{seats-votes election} if $\mathcal{V}_{P}$ is isomorphic to $\Delta_{P}$. Note that in such case, a vote redistribution function is simply a monotonic projection of the profile (i.e., vote share vector). We shall restrict our attention to a class of reasonable allocation functions we call, following the political science tradition, \emph{seats-votes functions} (though votes-seats functions would be more accurate).

\begin{definition}[Seats-Votes Function]\label{def:seatsVotes}
    A \emph{seats-votes function} is a function $f: \Delta \rightarrow \Delta$ fulfilling the following properties:
    \begin{description}[itemsep=2pt, parsep=0pt]

        \item[\textbf{symmetry}] $\sigma \circ f = f \circ \sigma$ for any $\sigma$ that permutes the coordinates of a point in $\Delta$,

        \item[\textbf{weak monotonicity}] $x_i > y_i$ and $x_j \le y_j$ for each $j \neq i$ implies $f_i(\mathbf{x}) \ge f_i(\mathbf{y})$ for every $\mathbf{x}, \mathbf{y} \in \Delta$ and $i \in \mathbb{N}$,

        \item[\textbf{negative unanimity}] $x_i = 0$ implies $f_i(\mathbf{x}) = 0$ for every $\mathbf{x} \in \Delta$ and $i \in \mathbb{N}$.

    \end{description}
\end{definition}

\noindent In other words, a seats-votes function maps a vector of vote shares to an election outcome (usually a vector of seat shares, but see the following remark). We require it to satisfy three natural axioms: The system should treat all parties in the same manner (\emph{symmetry}), getting more votes should always be non-detrimental (\emph{weak monotonicity}), and no party should be entitled to any seats without getting any votes (\emph{negative unanimity}). Note that the former two conditions have natural Arrovian counterparts.

\begin{remark}
    To note that the three axioms given above are independent, consider the following instances of seats-votes function that violate one while satisfying the others: a function with some positive bonus for party $1$ (which violates symmetry), a function with a guaranteed payoff for every party (which violates negative unanimity), and systems susceptible to no-show paradox \cite{Nurmi04,FelsenthalTideman13}, such as some variants of the largest remainders method using quota rounding \cite{Dancisin17}. 
\end{remark}

\begin{remark}
    Despite its name, a seats-votes function does not necessarily map vote shares to seat shares. For example, a function mapping voting weights to \emph{indices of voting power} 
    can also be formally considered as a seats-votes function under the above definition.
\end{remark}

\emph{Apportionment rules}, used to proportionally divide discrete goods like legislative seats \citep{BalinskiYoung01}, are the most common example of real-life electoral systems that can be described by seats-votes functions. We compare four common apportionment rules.\footnote{For the sake of simplicity and brevity, the definitions given here do not account for electoral ties and thus are irresolute for a subset of $\Delta_P$ (albeit one of Lebesgue measure zero) that corresponds to such ties. For a discussion of tie-breaking rules, see Appendix C.}

\begin{method}[First-Past-The-Post Rule]
    The \emph{FPTP apportionment rule}, also known as the \emph{plurality} or \emph{winner-takes-all} rule, is an allocation rule where the whole payoff is awarded to the largest party, i.e., $\mathbf{w} \in \Delta$ maps to a payoff vector $\mathbf{s} \in \Delta$ such that $s_i = 1$ only for $i$ satisfying $w_i = \max \mathbf{w}$, and all other coordinates are $0$.
\end{method}

\noindent Two subsequent methods are representatives of the class of \emph{divisor apportionment rules}:

\begin{definition}[Divisor Apportionment Rule \citep{Pukelsheim17}]
    For a fixed \emph{rounding rule} $r$, i.e., a non-decreasing function $r: \mathbb{R}_{\ge 0} \longrightarrow \mathbb{N}_{\ge 0}$ such that its restriction to $\mathbb{N}$ is an identity function, and a \emph{committee size} $k \in \mathbb{N}_{+}$, the divisor apportionment rule $\theta_{r}^{k}$ is an allocation rule that maps a vector $\mathbf{w} \in \Delta$ to a vector $(r(w_i \mu))_{i=1}^{\infty} / k$, where $\mu \in \mathbb{R}_{+}$ is such that $\lVert (r(w_i \mu))_{i=1}^{\infty} \rVert_{1} = k$.
\end{definition}

\noindent Intuitively, we seek to allocate $k$ seats in a committee by searching for a multiplier $M$ such that the sum of the products of $M$ and party vote shares rounded using $r$ equals $k$, and then allocating $r(w_i M)$ seats to the $i$-th party, $i = 1, \dots, n$. The examples we are interested in are:

\begin{method}[Jefferson--D'Hondt (JDH) Rule \citep{Jefferson92,DHondt82,BalinskiYoung78}]
    For a fixed {committee size} $k \in \mathbb{N}_{+}$, the \emph{Jefferson--D'Hondt rule} is the divisor rule obtained by rounding down, $\theta_{\lfloor\cdot\rfloor}^{k}$.
\end{method}

\begin{method}[Webster--Sainte-Laguë (WSL) Rule \citep{SainteLague10,BalinskiYoung75}]
    For a fixed {committee size} $k \in \mathbb{N}_{+}$, the \emph{WSL rule} is the divisor rule obtained by rounding to the nearest integer, $\theta_{\lfloor\cdot + 1/2\rfloor}^{k}$.
\end{method}

\noindent Intuitively, both Jefferson--D'Hondt and Webster--Sainte-Laguë methods are based on finding such a multiplier (divisor) that when all vote shares are multiplied (divided) by it and rounded down (in the case of JDH) or to the nearest integer (in the case of WSL), the resulting numbers of seats sum up to the desired committee size. The former approach is biased in favor of larger parties, while the latter is asymptotically unbiased.

Finally, we also consider the \emph{largest remainders rule}:

\begin{method}[Largest Remainders (LR) Rule \citep{BalinskiYoung75}]
    For a fixed {committee size} $k \in \mathbb{N}_{+}$, the \emph{largest remainders rule} is the allocation rule where each party's base payoff equals $\lfloor w_i k\rfloor / k$, and the unallocated payoff is divided equally among parties with $k - \lVert \lfloor w_i k\rfloor \rVert_{1}$ largest remainders $w_i k - \lfloor w_i k\rfloor$.
\end{method}

\noindent In other words, each party is initially awarded as many seats as its vote share multiplied by committee size and rounded down, and remaining seats are then awarded to parties with the largest remainders from the rounding-down operation. This rule is also asymptotically unbiased.

It is easy to see that, in the absence of ties, all of the methods described above are seats-votes functions.

Most real-life electoral systems are based on the above methods, except that the election is held in \emph{multiple electoral districts} and the allocation is performed in each district on the basis of the district-level rather than aggregate results. Common multi-district plurality systems like the ones used in, e.g., the United States, the United Kingdom, and Canada, afford well-known examples. Elections under such systems are not seats-votes elections, since the aggregate vote shares do not carry enough information to uniquely determine seat allocation, but they may still be approximated by seats-votes functions if we are willing to make some assumptions regarding the distribution of district-level vote share vectors \cite{Theil70,Gilliland85}.


\section{Spoiler-Proofness}\label{sec:proof}

Our main theoretical result here is an impossibility theorem regarding the existence of spoiler-proof seats-votes functions that are distinct from identity (i.e., perfect proportionality between votes and seats for each party).

\begin{theorem} \label{thm:nogo}
    The identity function given by $I(\mathbf{x}) = \mathbf{x}$ for $\mathbf{x} \in \Delta$ is the unique spoiler-proof seats-votes function.
\end{theorem}

\begin{proof}[Sketch of Proof]
     Let $f$ be a spoiler-proof seats-votes function. Fix any $i \in \mathbb{N}$ and $v \in (0, 1)$, and consider $\Lambda_{i} := \{ \mathbf{x} \in \Delta : x_i = v \}$. We begin by demonstrating that $f_i$ restricted to ${\Lambda_{i}}$ achieves maxima in the vertices of $\Lambda_{i}$. It is enough to show that spoiler-proofness implies that for any face of $\Lambda_{i}$ the function $f_i$ is maximized at its facets. Then we establish that $f_i \restriction_{\Lambda_{i}}$ also achieves minima in the vertices of $\Lambda_{i}$, by adding a virtual party $j$ and showing that every point in $\Lambda_{i}$ can be obtained by a monotonic projection from a vertex where $x_i = v, x_j = 1-v$.

    Accordingly, we arrive at the conclusion that $f_i(\mathbf{x})$ depends only on $x_i$, i.e., that there exists some $\varphi: [0, 1] \rightarrow [0, 1]$ such that $f_j(\mathbf{x}) = \varphi(x_j)$ for each $j \in \mathbb{N}$. From weak monotonicity and negative unanimity of $f$ we obtain a sequence of Schröder's functional equations $\varphi(px) = p \varphi(x)$ for any $x \in [0, \nicefrac{1}{p}]$. The proof that $\varphi(x) = x$ for all $x > 0$ is their unique solution follows the proof of Theorem A by \citet{TossavainenHaukkanen07}.

    For full proof, see Appendix A.
\end{proof}


To generalize this theorem beyond seats-votes elections, we need to introduce two further concepts: \emph{party independence of irrelevant alternatives} and \emph{zero-sum voting methods}. We first formalize the concept of two profiles being party-equivalent, i.e., differing only as to the distribution of votes among parties other than some fixed party $i$. To simplify notation, for ranked voting methods we denote the rank position of $i$ in $\mathbf{v} \in \mathcal{D}_{P}$ by $\mathbf{v}_i := |k \in P: k \preceq_{\mathbf{v}} i|$.

\begin{definition}[Party-Equivalence of Profiles]
    Fix any $P \in \mathcal{P}(\mathbb{N})$ and any $i \in P$. Two profiles $\mathbf{x}, \mathbf{y} \in \mathcal{V}_{P}$ are $i$-equivalent if and only if there exists a bijective map $\eta$ between votes in $\mathbf{x}$ and in $\mathbf{y}$ such that $\mathbf{v}_i = \mathbf{u}_i$ for every $(\mathbf{v}, \mathbf{u}) \in \eta$.
\end{definition}

Intuitively, an electoral rule satisfies party independence of irrelevant alternatives if every party's allocation is independent from the distribution of votes among other parties, or, in other words, constant for all party-equivalent profiles:

\begin{definition}[Party Independence of Irrelevant Alternatives]
    An electoral rule $f$ satisfies \textbf{party independence of irrelevant alternatives} is for every $P \in \mathcal{P}(\mathbb{N})$, every party $i \in P$, and every pair of profiles $\mathbf{x}, \mathbf{y} \in \mathcal{V}_{P}$ that are $i$-equivalent $f(\mathbf{x})_i = f(\mathbf{y})_i$.
\end{definition}

Second, zero-sum voting methods are such where a voter cannot improve the score or rank of one party without worsening the score or rank of another.

\begin{definition}[Zero-Sum Voting Method]
    A voting method is \emph{zero-sum} if, for any $x, y \in \mathcal{D}_{P}$, $\mathbf{x}_i > \mathbf{y}_i$ for some $i \in P$ implies $\mathbf{x}_j < \mathbf{y}_j$ for some $j \in P$.
\end{definition}

\noindent Examples of zero-sum voting methods include single-choice, cumulative, ordinal, and knapsack voting, while examples of non-zero sum voting methods include approval and scoring.

Finally, an allocation rule is \emph{negative-unanimous} if and only if each party $i$ such that for every vote $\mathbf{v} \in \mathbf{w}$ there exists no vote $\mathbf{u} \in \mathcal{D}_P$ satisfying $\mathbf{u} \prec_{i} \mathbf{v}$ obtains a zero payoff.

\begin{theorem} \label{thm:nogoGeneral}
    For any zero-sum voting method a neutral, anonymous, weakly monotonic, and negative-unanimous allocation rule $f$ is spoiler-proof if and only if it satisfies party independence of irrelevant alternatives.
\end{theorem}

\noindent The proof of the above theorem mirrors the first part of the proof of Theorem \ref{thm:nogo}. See Appendix B for details.

Note that Theorem \ref{thm:nogoGeneral} is a generalization of earlier findings regarding spoilers in single-winner elections: ordinal voting is zero-sum, and monotonic ordinal voting rules fail to satisfy IIA \cite{Arrow50}, hence it is not spoiler-proof \cite{McCuneWilson24}, but approval voting is not zero-sum and accordingly spoiler-proof (assuming sincere voting) \cite{BrandlPeters22}.

\section{Spoiler Susceptibility}\label{sec:experiments}

The natural next step is to consider whether spoilers are low-probability outliers or something that arises regularly, as well as how their probability and magnitude depend on the electoral rule. 
Our motivation is that even if we cannot eliminate spoilers, perhaps we can find rules for which average spoiler impact will be minimized. To formalize that idea, we introduce the concept of \emph{spoiler susceptibility}.


How to go from the definition of an individual spoiler's impact (Definition \ref{def:spoilership}) to a definition of spoiler susceptibility? First, fix a family of vote redistribution functions, $(\rho_i)_{i \in P}$. We define spoiler susceptibility of an allocation rule $f$ as the expected maximum excess electoral impact, where the maximum is taken for each election
over parties, and the expectation is then taken over some probability distribution, $\mathcal{D}$, on the set of profiles, $\mathcal{V}_{P}$:
\begin{equation}
    \Phi(\mathcal{D}, f) := \mathbb{E} \, \max \limits_{i \in P} \lambda_{i, \rho_i}^f (\mathbf{W}), \textrm{ where } \mathbf{W} \sim \mathcal{D},
\end{equation}
where $\mathbf{W}$ is the random profile. As spoiler susceptibility is difficult to model analytically\footnotemark, we study it experimentally. We start by describing probabilistic models for generating elections.
\footnotetext{The difficulty stems primarily from realistic vote distributions rarely admitting closed-form integrals. For an example, see results for the probability of occurrence of spoilers under the Jefferson--D'Hondt rule in \citep{BoratynEtAl25}.}

\subsection{Probabilistic Models}

The probabilistic models commonly used in computational social choice \citep{BoehmerEtAl24} have been developed for single-district elections. However, as noted in Sec. \ref{sec:seatsVotes}, most of the real-life party elections are held in multiple districts. We cannot simulate a multi-district election by simply generating multiple instances of one election, since real-life districts are not unbiased samples from the population of voters.\footnote{Imagine they were. Then, per the central limit theorem, as the number of voters increases, the outcomes in each district would converge to the outcome for the population (commingled) profile. For instance, if we were to consider the FPTP rule, the same party would win in every district and obtain all the seats. Clearly, this is not what happens in real-life FPTP elections.}

We consider four classes of probabilistic models:
\smallskip

\noindent\textbf{Spatial Models.}
    In a $d$-dimensional Euclidean model parties and voters are assigned ideal points in $\mathbb{R}^{d}$ \citep{EnelowHinich84,Merrill84}. First, party positions are drawn from $\operatorname{Unif}((0, 1)^{d})$, and then candidate positions (one per district from each party) are drawn from the multivariate normal distribution centered at the party's ideal point with covariance matrix $\mathbf{\Sigma} := \sigma I_{d}$, where $I_{d}$ is a $d \times d$ identity matrix and $\sigma \in \mathbb{R}_{+}$. Voter ideal points are drawn independently from the uniform distribution on $(0, 1)^{d}$, then shifted in each district independently by a vector drawn from $\operatorname{Unif}(-1/4, 1/4)^{d}$.
\smallskip

\noindent\textbf{Single-Peaked Models.}
    We consider two models for generating single-peaked profiles. In one \citep{wal:t:generate-sp} we are given an ordering on the set of candidates, and each vote is drawn from the uniform distribution on all single-peaked votes consistent therewith. In the other one \citep{con:j:eliciting-singlepeaked} the peak is drawn from a uniform distribution on candidates, and the rest of the vote is obtained by a random walk.
\smallskip

\noindent\textbf{Mallows Model.}
    The Mallows model \citep{Mallows57,CritchlowEtAl91} is parametrized by a single parameter $\phi \in [0,1]$, and a (central) vote $v_c \in \mathcal{D}_{P}$. The probability of generating a vote $v$ is proportional to $\phi^{\kappa(v_c, v)}$, where $\kappa(v_c, v)$ is the Kendall tau distance \citep{Kendall38} between $v_c$ and $v$, i.e., the minimum number of swaps of adjacent candidates needed to transform the vote $v$ into the central vote $v_c$.
    We first generate a central vote for each district, $v_c^i$ with parameter $\phi_{1}$ and a starting vote $v_c^0 := [p]$, then generate votes within each district with parameter $\phi_{2}$ and a district--wide central vote $v_c^i$. On sampling, see \citet{LuBoutilier14}.\footnote{We use a Mallows model parameterization by \citet{BoehmerEtAl21}, based on a normalized dispersion parameter $\normphi$.}
\smallskip

\noindent\textbf{Impartial Culture (IC).}
    Under IC, each vote is drawn randomly from the uniform distribution on linear orders on candidates \citep{CampbellTullock65}. This model does not account for intra-district clustering, and is regarded as a poor approximation of real-life instances \citep{RegenwetterEtAl06,TidemanPlassmann12}.

Because all of those cultures yield full preference orderings, there is no need to separately model redistributions. Instead, a unique redistribution function maps each preference ordering to an ordering on the set of remaining parties.

\subsection{Allocation Rules}

We analyze four allocation rules discussed in Sec. \ref{sec:seatsVotes}: FPTP, Jefferson--D'Hondt, Webster-Sainte-Laguë, and Largest Remainders. In each of those rules, seats are independently apportioned among parties in $256$ equal-size electoral districts on the basis of party vote shares in a given district. For each rule, we consider two definitions of vote shares. Under one variant, we define them as the number of votes where a given party is ranked first, and in the other one, as the sum of Borda scores over such candidates. In both cases, the vote shares are normalized so as to add up to $1$.

\subsection{Experimental Results}
To compare voting methods, we analyze their performance under $8$ probabilistic models: four spatial models ($d = 1, 2$ and $\sigma = 0.05, 0.2$), two single-peaked models (Walsh \citep{wal:t:generate-sp} and Conitzer \citep{con:j:eliciting-singlepeaked}), one Mallows model ($\phi_1 =\phi_2 = 0.75$) and impartial culture. For every model, committee size $k \in \{5, 10, 15\}$, and number of parties $p \in \{3,\dots,12\}$ we run $300$ simulations, with the number of districts and the number of voters per district both equal to $256$. The figures for $k = 5$ are reported above (others can be found in Appendix E).

Experimental results demonstrate several regularities. First, and most importantly, spoiler susceptibility strongly correlates with the deviation from perfect proportionality, whether measured by the $L_2$ distance between the vote share and its image under the allocation rule, or by the $\alpha$-divergence from the latter to the former (for $\alpha=1$ and $\alpha = 0$) \citep{Wada10}. Thus, Webster--Sainte-Laguë is least spoiler susceptible (and most proportional), followed by Largest Remainders, Jefferson--D'Hondt, and finally FPTP (with the last one being an outlier).
Moreover, for all rules but FPTP, increasing committee size $k$ causes both disproportionality and spoiler susceptibility to decrease.

Second, switching from first-choice apportionment to apportionment on the basis of Borda scores substantially decreases spoiler susceptibility and virtually eliminates the differences between WSL, JDH, and LR. Finally, spoiler susceptibility generally increases with the number of parties, although there are exceptions to this rule (e.g., the Walsh model). It is also worth noting that the spoiler problem is particularly acute under the Conitzer and Walsh models. For Conitzer, spoiler susceptibility of the FPTP rule can reach as high as $1.2$, meaning that more than half of the total payoff is redistributed.

\setlength{\abovecaptionskip}{3pt}
\setlength{\belowcaptionskip}{6pt}

\begin{figure}[t]
    \begin{tikzpicture}
        \node (img)  { \includegraphics[width=\columnwidth]{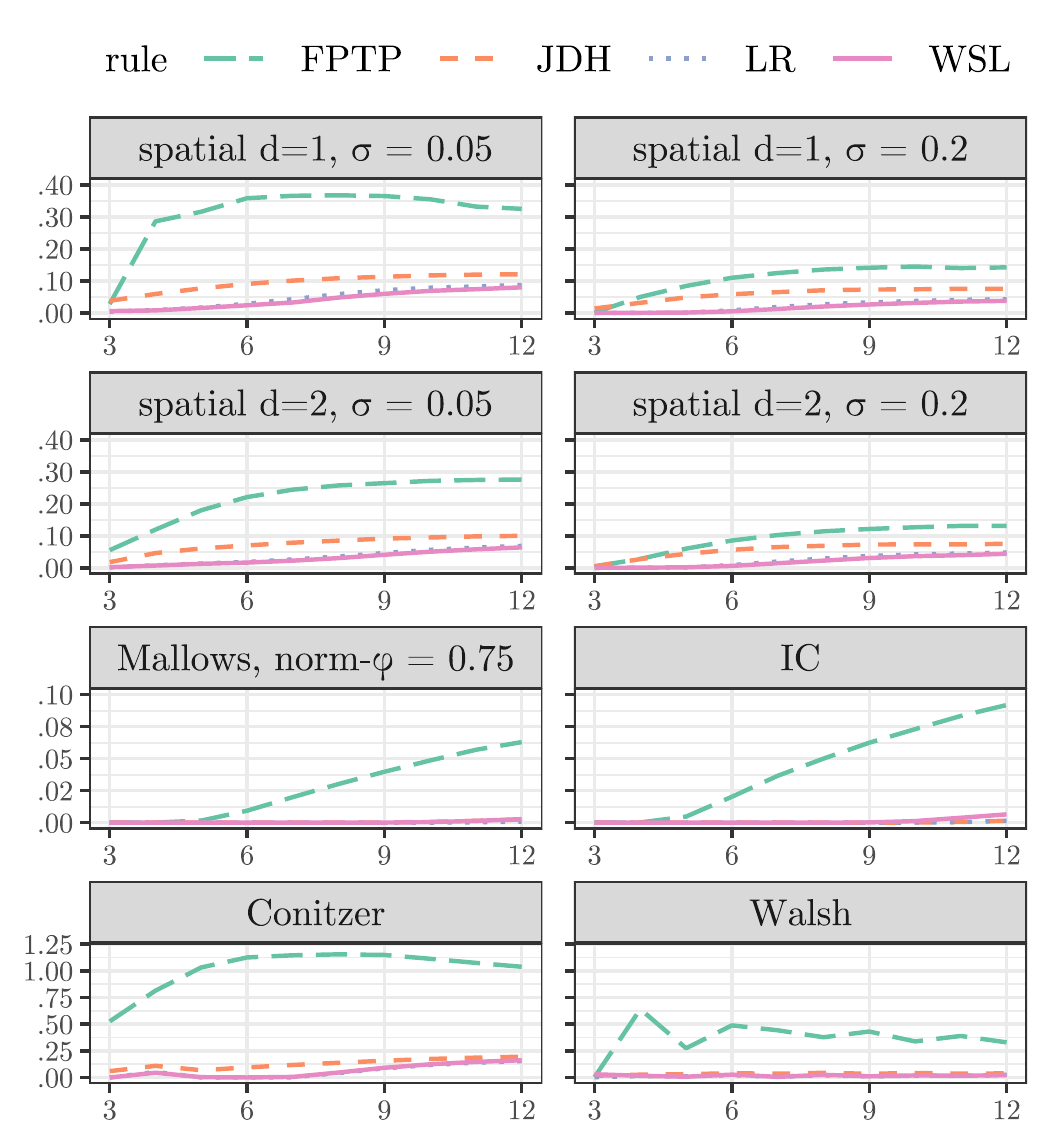}};
        \node[below=of img, node distance=0cm, yshift=1cm] {$p$ (number of parties)};
        \node[left=of img, node distance=0cm, rotate=90, anchor=center,yshift=-1.1cm] {$\Phi$ (spoiler susceptibility)};
    \end{tikzpicture}
    \caption{Spoiler susceptibility of electoral rules -- apportionment on the basis of first preferences. Committee size, $k = 5$, is constant in each district. Note that the vertical scale differs between rows.}
    \label{fig:results}
\end{figure}

\begin{figure}[t]
    \begin{tikzpicture}
        \node (img)  { \includegraphics[width=\columnwidth]{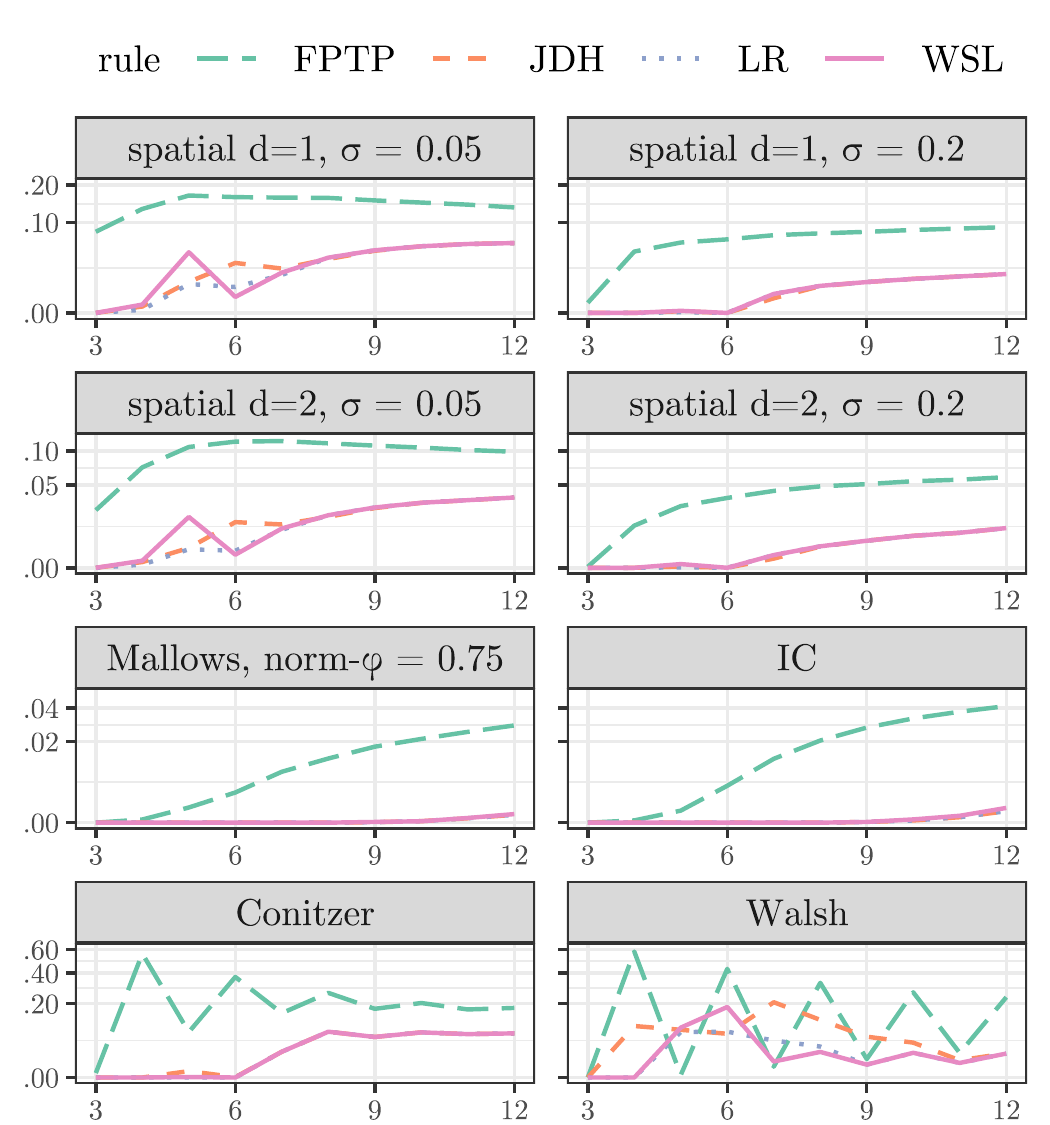}};
        \node[below=of img, node distance=0cm, yshift=1cm] {$p$ (number of parties)};
        \node[left=of img, node distance=0cm, rotate=90, anchor=center,yshift=-1.1cm] {$\Phi$ (spoiler susceptibility)};
    \end{tikzpicture}
    \caption{Spoiler susceptibility of electoral rules -- apportionment on the basis of Borda scores. Committee size, $k = 5$, is constant in each district. The vertical scale is logarithmic and differs between rows.}
    \label{fig:resultsBorda}
\end{figure}

\section{Conclusions}
We have introduced a novel approach to define spoilers in party elections that yields both theoretical and experimental results. In particular, we show that spoiler-proofness is a very strong postulate: The identity function, corresponding to proportional allocation of seats, is the unique spoiler-proof seats-votes function, and for zero-sum voting methods in general spoiler-proofness is equivalent to party independence of irrelevant alternatives, which (like the original Arrovian IIA) is a very strong condition.

We have also introduced a new, continuous measure of spoiler impact, noting that spoiler susceptibility is not binary. We find that the magnitude of spoiler impact varies with the allocation rule. More proportional rules like Webster or Largest Remainders are less susceptible to spoiler effects, while -- as expected -- FPTP (winner-takes-all) exhibits particularly high spoiler susceptibility. Moreover, spoiler resilience is enhanced by increasing district magnitude (per-district committee size) and basing apportionment on Borda scores instead of just the first choices.

Future work will focus on relaxations of spoiler-proofness, in particular on approximate spoiler-proofness. Here, we conjecture that the relation between proportionality, measured by $L_1$ distance, and spoiler susceptibility is $1$-Lipschitz-continuous.

Future work will also address spoiler susceptibility of non-seats votes elections. Our experimental techniques can be particularly easily applied to ordinal elections in general, making it possible to test the claims of non-seats-votes methods, such as STV, to superior performance insofar as spoiler resilience is concerned. Another area that will be explored is spoiler susceptibility of common participatory budgeting rules and approval party elections in general.

\section{Acknowledgements}

This research has been supported under Polish National Center for Science grant no. 2019/35/B/HS5/03949 and by the French government under the management of Agence Nationale de la Recherche as part of the France 2030 program, reference ANR-23-IACL-0008.

\newpage

\renewcommand\thesubsection{\Alph{subsection}}
\section*{Appendix}

\subsection{Proof of Theorem \ref{thm:nogo}}

\begin{proof}
    Let $f$ be a spoiler-proof seats-votes function. Fix $i \in P$ and $v \in (0, 1)$, and consider $\Lambda_{i} := \{ \mathbf{x} \in \Delta_{P} : x_i = v \}$. We begin by demonstrating that $f_i {\restriction}_{\Lambda_{i}}$ achieves maxima in the vertices of $\Lambda_{i}$.
    
    It is enough to show that for an arbitrary face of $\Lambda_{i}$ the function $f_i$ is maximized at its facets. 
    Each face of $\Lambda_i$ is a set of form $L_K := \{ \mathbf{x} \in \Lambda_i : \sum_{k \in K} x_k = 1 - v \}$ for some nonempty $K \subseteq P \setminus \{i\}$, with $x_l = 0$ for $l \in P \setminus (\{i\} \cup K)$. Let us denote its relative interior, i.e., the set of points $\mathbf{x}$ such that $x_k > 0$ for every $k \in K$, by $\relint L_K$, and its relative boundary, i.e., the union of its facets, by $\relbd L_K$. Assume on the contrary that there exists some $\mathbf{y} \in \relint{L_K}$ such that $f_i(\mathbf{y}) > \sup_{\mathbf{x} \in \relbd L_K} f_i(\mathbf{x})$. But from the spoiler-proofness of $f$, for every $j \in K$ and $\pi \in \Pi_{P,-j}$ we have $f_i(\pi(\mathbf{y})) \ge f_i(\mathbf{y})$. From the arbitrariness of the choice of $\pi$ we can assume that $\pi(L_K) \subset L_K$, i.e., $\pi_l(\mathbf{x}) = x_l$ for every $l \in P \setminus K$. Then, $\pi(\relint{L_K}) \subset \relbd L_K$, and we arrive at a contradiction.
    
    We can apply this reasoning inductively until we arrive at $\dim L_K = 0$. From the symmetry of $f$ it follows that $f_i$ is equal for every vertex of $\Lambda_i$.

    On the other hand, we will now show that $f_i \restriction_{\Lambda_{i}}$ also achieves minima in the vertices of $\Lambda_{i}$. Let us first consider $\mathbf{z} \in \Delta_{P'}$, where $P' := P \, \cup \, \{q\}$, where $q$ is the smallest ordinal not in $P$, such that $z_i = v$, $z_q = 1 - v$, and $z_j = 0$ for $j \in P \setminus \{i\}$. Note that for every $\mathbf{x} \in \Lambda_i$ there exists some projection $\pi \in \Pi_{P',-q}$ such that $\pi(\mathbf{z}) = \mathbf{x}$. From the spoiler-proofness of $f$ it follows that $f_i(\mathbf{z}) \le \inf_{\mathbf{x} \in \Lambda_{i}} f_i(\mathbf{x}) \le \inf_{\mathbf{x} \in \relbd \Lambda_{i}} f_i(\mathbf{x})$. But from the symmetry of $f$ we have $f_i(\mathbf{z}) = f_i(\mathbf{z}^{j})$ for every $j \in P \setminus \{i\}$, $\mathbf{z}^{j} \in \Delta_{P'}$, defined as $z_i^j := v$, $z_j^j := 1 - v$, and $z_k^j := 0$ for $k \in P' \setminus \{i, j\}$. As the set $\{\mathbf{z}^{j} : j \in P \setminus \{i\}\}$ is the set of vertices of $\Lambda_{i}$, we conclude that $f_i \restriction_{\Lambda_{i}}$ achieves minima in them. Thus, $f_i$ is constant over $\Lambda_{i}$.

    Accordingly, we arrive at the conclusion that $f_i(\mathbf{x})$ is independent of all coordinates of $\mathbf{x}$ except $x_i$. From the arbitrariness of the choice of $i$, the same holds for every other coordinate. Thus, by symmetry of $f$, there exists a function $\varphi: [0, 1] \rightarrow [0, 1]$ such that $f_i(\mathbf{x}) = \varphi(x_i)$ for every $i \in P$. From weak monotonicity of $f$ it follows that $\varphi$ is non-decreasing, while from negative unanimity it follows that $\varphi(0) = 0$ and $\varphi(1) = 1$. It also follows that for every $\mathbf{x} \in \Delta_{k}$, $k \in \mathbb{N}_{+}$, we have $\sum_{i=1}^{k} \varphi(x_i) = 1$. Hence, in particular, for every $n \in \mathbb{N}_{+}$ and every $x \in [0, 1/n]$ we have $n \varphi(x) + \varphi(1 - nx) = 1$. On the other hand, $\varphi(1-y) = 1 - \varphi(y)$ for every $y \in [0, 1]$. Thus, we obtain a sequence of Schröder's functional equations $\varphi(nx) = n \varphi(x)$ for every $x \in [0, 1/n]$, $n \ge 2$.

    Fix any $x \in \mathbb{Q} \, \cap \, (0, 1)$. There exist $a < b \in \mathbb{N}_{+}$ such that $x = a / b$. As $1/b \in [0, 1/a]$, it follows that $\varphi(a/b) = a \varphi(1/b)$. However, as $1/b \in [0, 1/b]$, it also follows that $b \varphi(1/b) = \varphi(b/b) = 1$, and thus $\varphi(x) = \varphi(a/b) = a/b = x$ for every $x \in \mathbb{Q} \cap [0, 1]$. It remains for us to show that the same holds for every $x \in [0, 1] \setminus \mathbb{Q}$. Fix any such $x$ and take rational sequences $(r_n)_{n \in \mathbb{N}_{+}} \uparrow x$ and $(s_n)_{n \in \mathbb{N}_{+}} \downarrow x$. From the monotonicity of $\varphi$ it follows that $r_n = f(r_n) \le f(x) \le f(s_n) = s_n$ for every $n \in \mathbb{N}_{+}$, and accordingly that $f(x) = x$ as $n \rightarrow \infty$.

\end{proof}

\subsection{Proof of Theorem \ref{thm:nogoGeneral}}\label{sec:proofNogoGeneral}

\begin{definition}
    For a ranked voting method a party $i \in P$ is minimal under a vote $\mathbf{v} \in \mathcal{D}_{P}$ if $i \preceq_{\mathbf{w}} j$ for every $j \in P$. \smallskip \\
    For a cardinal voting method a party $i \in P$ is minimal under a vote $\mathbf{v} \in \mathcal{D}_{P}$ if $\mathbf{v}_{i} \le \mathbf{u}_{i}$ for every vote $\mathbf{u} \in \mathcal{D}_{P}$.
\end{definition}

\begin{definition}
    Fix any $i \in P$ and any equivalence class of profiles with respect to $i$-equivalence, $\Lambda_{i} \in \mathcal{V}_{P} \big/ {\sim_{i}}$. A $K$-face of $\Lambda$, $K \subseteq P$, is any set of profiles $V \in \Lambda$ that $i$ is minimal under $\mathbf{v}$ for every $i \in P \setminus K$ and every $\mathbf{v} \in \mathbf{w} \in V$. A facet of $\Lambda$ is any set of profiles $V \in \Lambda$ such that $i$ is minimal under $\mathbf{v}$ for every $\mathbf{v} \in \mathbf{w} \in V$ for exactly one $i \in \{j \in P: \exists x, y \in \Lambda: x \prec_j y\}$.
\end{definition}

\begin{proof}
    Consider first the case of \textbf{cardinal voting rules}:

    Let $f$ be a spoiler-proof allocation rule. Fix any $i \in P$ and any equivalence class of profiles with respect to $i$-equivalence, $\Lambda_{i} \in \mathcal{V}_{P} \big/ {\sim_{i}}$. We begin by demonstrating that $f_i \restriction_{\Lambda_{i}}$ achieves maxima in the vertices of $\Lambda_{i}$, i.e., such profiles that are minimal for all parties except $i$ and some $j \in P \setminus {i}$. It trivially follows that $\Lambda_i$ has at most $|P \setminus \{i\}|$.

    It is enough to show that for any face of $\Lambda_{i}$ the function $f_i$ is maximized at its facets. Let $L_K$ be a $K$-face of $\Lambda_{i}$ for some non-empty $K \subseteq P \setminus \{i\}$. Let us denote its relative interior, i.e., the set of profiles $\mathbf{x}$ such that no $k \in K$ is minimal under $\mathbf{x}$, by $\relint L_K$, and its relative boundary, i.e., the union of its facets, by $\relbd L_K$. Assume on the contrary that there exists some $Y \in \relint{L_K}$ such that $f_i(Y) > \sup_{X \in \relbd L_K} f_i(X)$, where $\relbd L_K$ denotes the boundary of $L_K$, i.e., the union of its facets. But from the spoiler-proofness of $f$, it follows that $f_i(\pi(Y)) \ge f_i(Y)$ for every $j \in K$ and $\pi \in R_{-j}$. From the arbitrariness of the choice of $\pi$ we can assume that $\pi(L_K) \subset L_K$, i.e., $\pi_l(\mathbf{x}) \equiv_l \mathbf{x}$ for every $l \in P \setminus K$. Then, $\pi(\relint{L_K}) \subset \relbd L_K$, and we arrive at a contradiction.

    We can apply this reasoning inductively until facets of $L_K$ become vertices of $\Lambda_i$. From the symmetry of $f$ it follows that $f_i$ is equal for every vertex of $\Lambda_i$.

    On the other hand, we will now show that $f_i \restriction_{\Lambda_{i}}$ also achieves minima in the vertices of $\Lambda_{i}$. Fix any $k \in \mathbb{N} \setminus P$. Let $P' = P \cup \{k\}$, and consider $\mathbf{z} \in \mathcal{V}_{P'}$ be such that $\mathbf{z} \equiv_i \mathbf{x}$ for any $\mathbf{x} \in \Lambda_{i}$ and every $j \in P \setminus \{i\}$ is minimal under $\mathbf{z}$. Note that for every $\mathbf{x} \in \Lambda_i$ there exists some vote redistribution function $\pi$ such that $\pi(\mathbf{z}) = \mathbf{x}$. From the spoiler-proofness of $f$ it follows that $f_i(\mathbf{z}) \le \inf_{\mathbf{x} \in \partial \Lambda_{j}} f_i(\mathbf{x})$. But from the symmetry of $f$ we have $f_i(\mathbf{z}) = f_i(\mathbf{z}^{j})$ for every $\mathbf{z}^{j} \in \mathcal{D}_{P'}$, where $\mathbf{z}^{j}$, $j \in P \setminus \{i\}$, are the vertices ($1$-faces) of $[\mathbf{z}]_{\equiv_i}$ other than $\mathbf{z}$. 
    As the set $\{\mathbf{z}^{j} : j \in P \setminus \{i\}\}$ is the set of vertices of $\Lambda_{i}$, we conclude that $f_i \restriction_{\Lambda_{i}}$ achieves minima there. Thus, $f_i$ is constant over $\Lambda$, as desired.

    The proof in the other direction trivially follows from symmetry and weak monotonicity of $f$.

\end{proof}

\subsection{Tie-Breaking for Apportionment Rules}\label{app:tieBreak}

\subsubsection{Jefferson--D'Hondt and Webster--Sainte-Laguë}

An \emph{electoral tie} under Jefferson--D'Hondt and Webster--Sainte-Laguë occurs if and only if there exists no multiplier satisfying $\lVert (r(w_i M))_{i=1}^{\infty} \rVert_{1} = k$, where $r(\cdot)$ denotes $\lfloor \cdot \rfloor$ or $[\cdot]$, as the case may be. Intuitively, our tie-breaking rule is to take an expectation over random resolutions. Formally, the rule is as follows:
\begin{itemize}
\item Let $M_{-}$ be the largest such $M \in \mathbb{R}_{+}$ such that $\lVert (r(w_i M))_{i=1}^{\infty} \rVert_{1} < k$, and let $M_{+}$ be the largest such $M \in \mathbb{R}_{+}$ such that $\lVert (r(w_i M))_{i=1}^{\infty} \rVert_{1} > k$.
\item Let $\tau := (r(w_i M_{+}))_{i=1}^{\infty} - (r(w_i M_{-}))_{i=1}^{\infty}$.
\item Let $f(\mathbf{w}) := ((r(w_i M_{-}))_{i=1}^{\infty} + \tau / \lVert\tau\rVert_{1}) / k$
\end{itemize}

\subsubsection{Largest Remainders}

An \emph{electoral tie} under Largest Remainders occurs if $\sum_{i=1}^{\infty} r_i(\mathbf{w}, k) > k - \lVert \lfloor w_i k\rfloor \rVert_{1}$. Again, our tie-breaking rule is to take an expectation over random resolutions. Formally, it is as follows:
\begin{itemize}
\item Let $\tau$ be the number of remainders, $\{w_i k\}$, equal to the $k$-th largest such remainder.
\item Let $r_i^{*}(\mathbf{w}, k)$ equal $1$ iff $\{w_i k\}$ is greater than the $k$-th largest remainder, $\tau^{-1}$ iff $\{w_i k\}$ equals the $k$-th largest remainder, and $0$ otherwise.
\item Let $f(\mathbf{w}) := (\lfloor w_i k \rfloor + r_i^{*}(\mathbf{w}, k))_{i=1}^{\infty} / k$.
\end{itemize}

\subsubsection{FPTP}

An \emph{electoral tie} under FPTP occurs if there are several parties $i_1, \dots, i_k \in P$ such that $v_{i_j} = \max_{\ell = 1, \dots, n} v_{\ell}$ for each $j = 1, \dots, k$. In such a case, the payoff is equally divided among those parties, i.e., $s_{i_j} = 1/k$ for every $j = 1, \dots, k$.

\FloatBarrier
\subsection{Experimental Results for $k = 10, 15$.}\label{sec:appExperimental}

\begin{figure}[b]
    \begin{tikzpicture}
        \node (img)  { \includegraphics[width=.98\columnwidth]{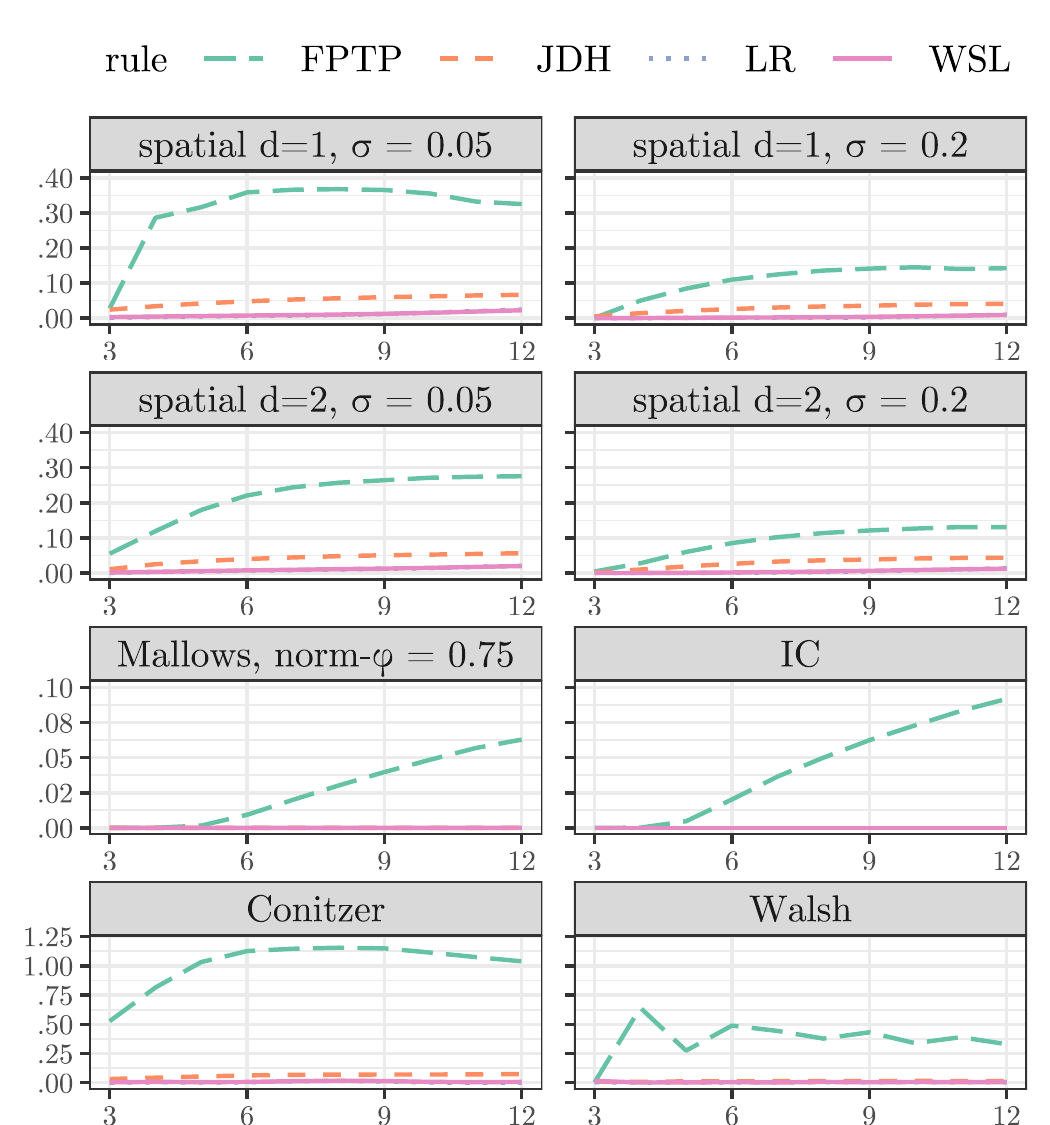}};
        \node[below=of img, node distance=0cm, yshift=1cm] {$p$ (number of parties)};
        \node[left=of img, node distance=0cm, rotate=90, anchor=center,yshift=-1.1cm] {$\Phi$ (spoiler susceptibility)};
    \end{tikzpicture}
    \caption{Spoiler susceptibility of electoral rules -- apportionment on the basis of first preferences. Committee size, $k = 10$, is constant in each district. Note that the vertical scale differs between rows.}
    \label{fig:results10}
\end{figure}

\begin{figure}[t]
    \begin{tikzpicture}
        \node (img)  { \includegraphics[width=.98\columnwidth]{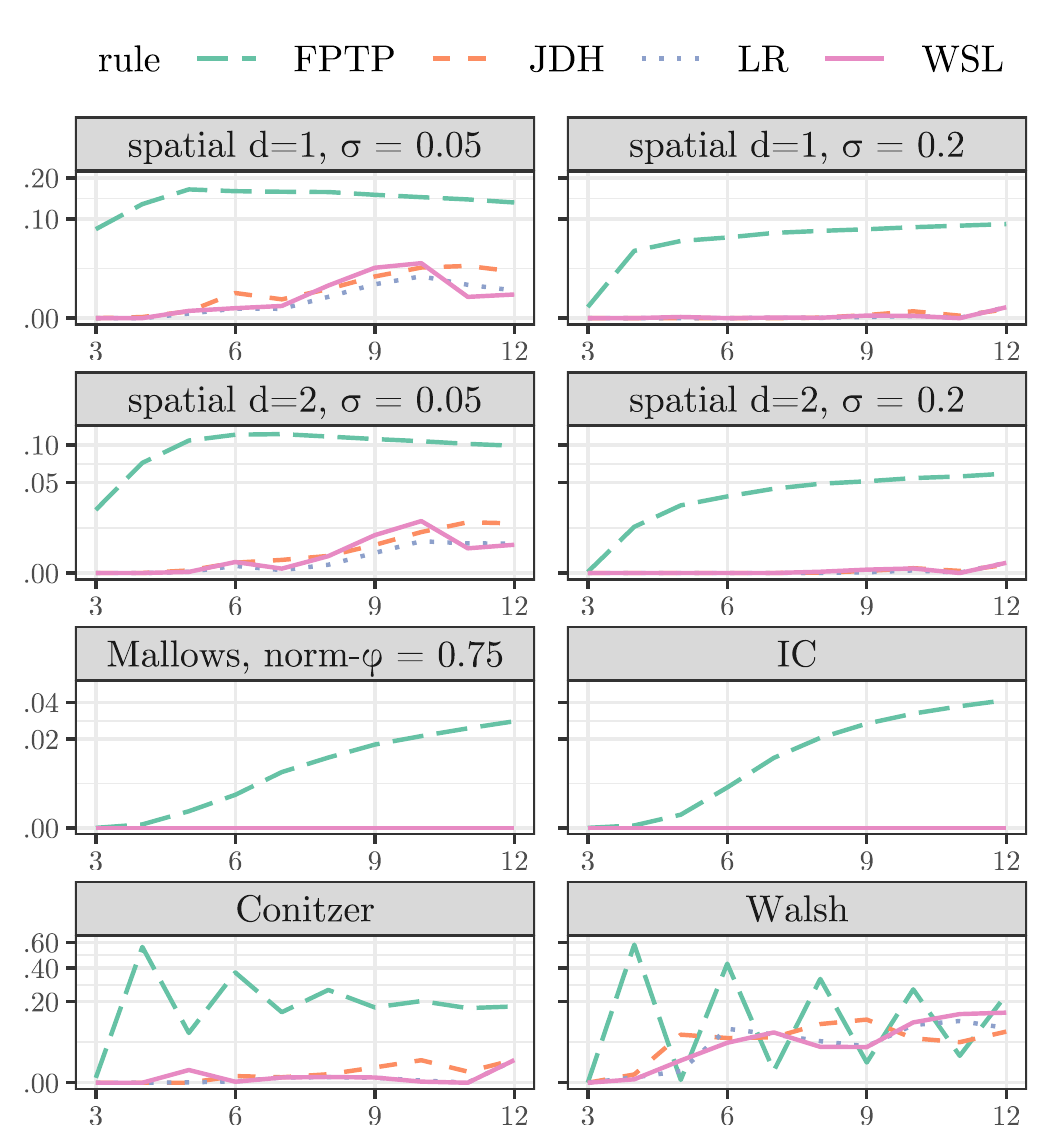}};
        \node[below=of img, node distance=0cm, yshift=1cm] {$p$ (number of parties)};
        \node[left=of img, node distance=0cm, rotate=90, anchor=center,yshift=-1.1cm] {$\Phi$ (spoiler susceptibility)};
    \end{tikzpicture}
    \caption{Spoiler susceptibility of electoral rules -- apportionment on the basis of Borda scores. Committee size, $k = 10$, is constant in each district. Note that the vertical scale differs between rows.}
    \label{fig:results10b}
\end{figure}

\begin{figure}[t]
    \begin{tikzpicture}
        \node (img)  { \includegraphics[width=.98\columnwidth]{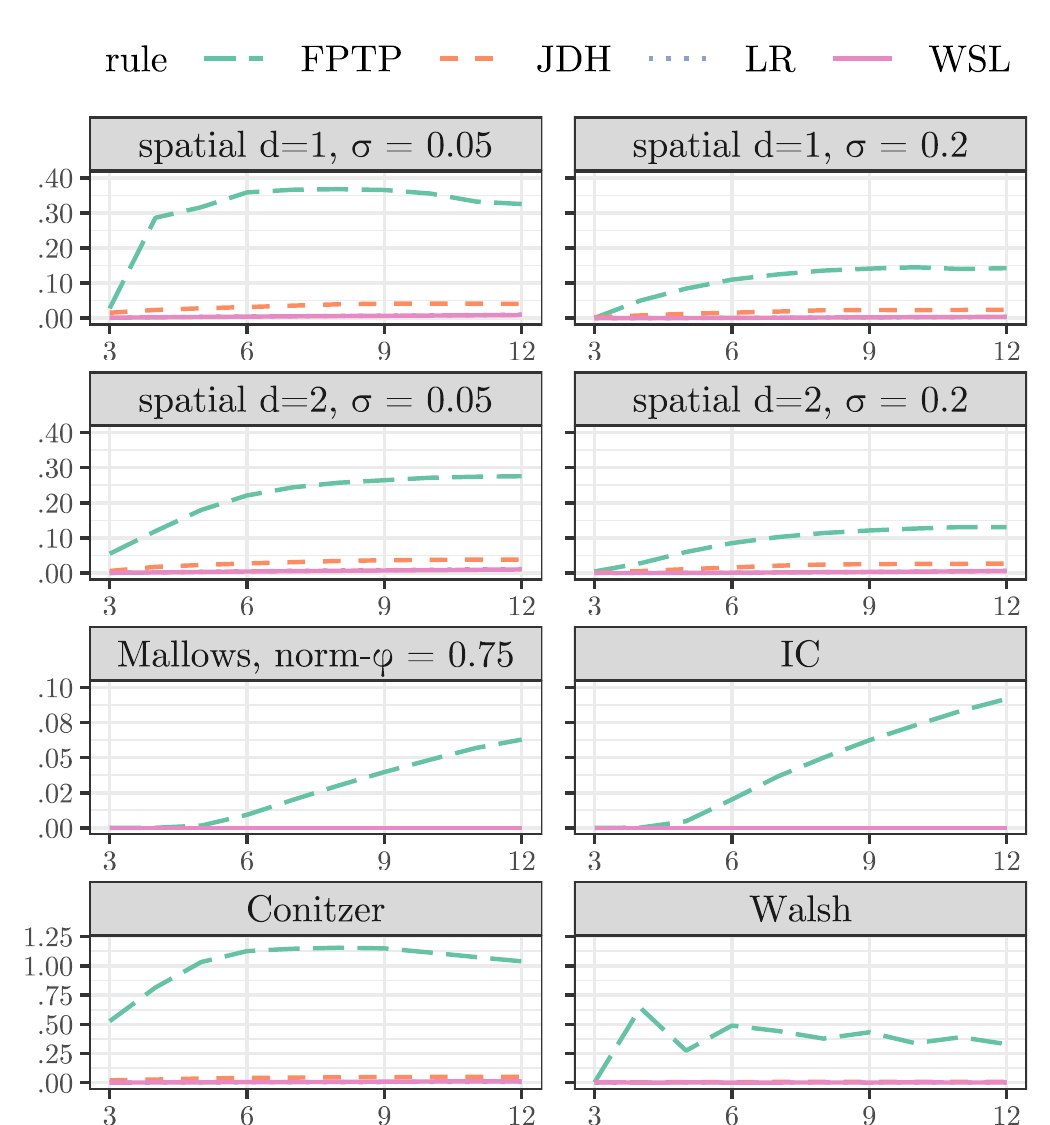}};
        \node[below=of img, node distance=0cm, yshift=1cm] {$p$ (number of parties)};
        \node[left=of img, node distance=0cm, rotate=90, anchor=center,yshift=-1.1cm] {$\Phi$ (spoiler susceptibility)};
    \end{tikzpicture}
    \caption{Spoiler susceptibility of electoral rules -- apportionment on the basis of first preferences. Committee size, $k = 15$, is constant in each district. Note that the vertical scale differs between rows.}
    \label{fig:results15}
\end{figure}

\begin{figure}[t]
    \begin{tikzpicture}
        \node (img)  { \includegraphics[width=.98\columnwidth]{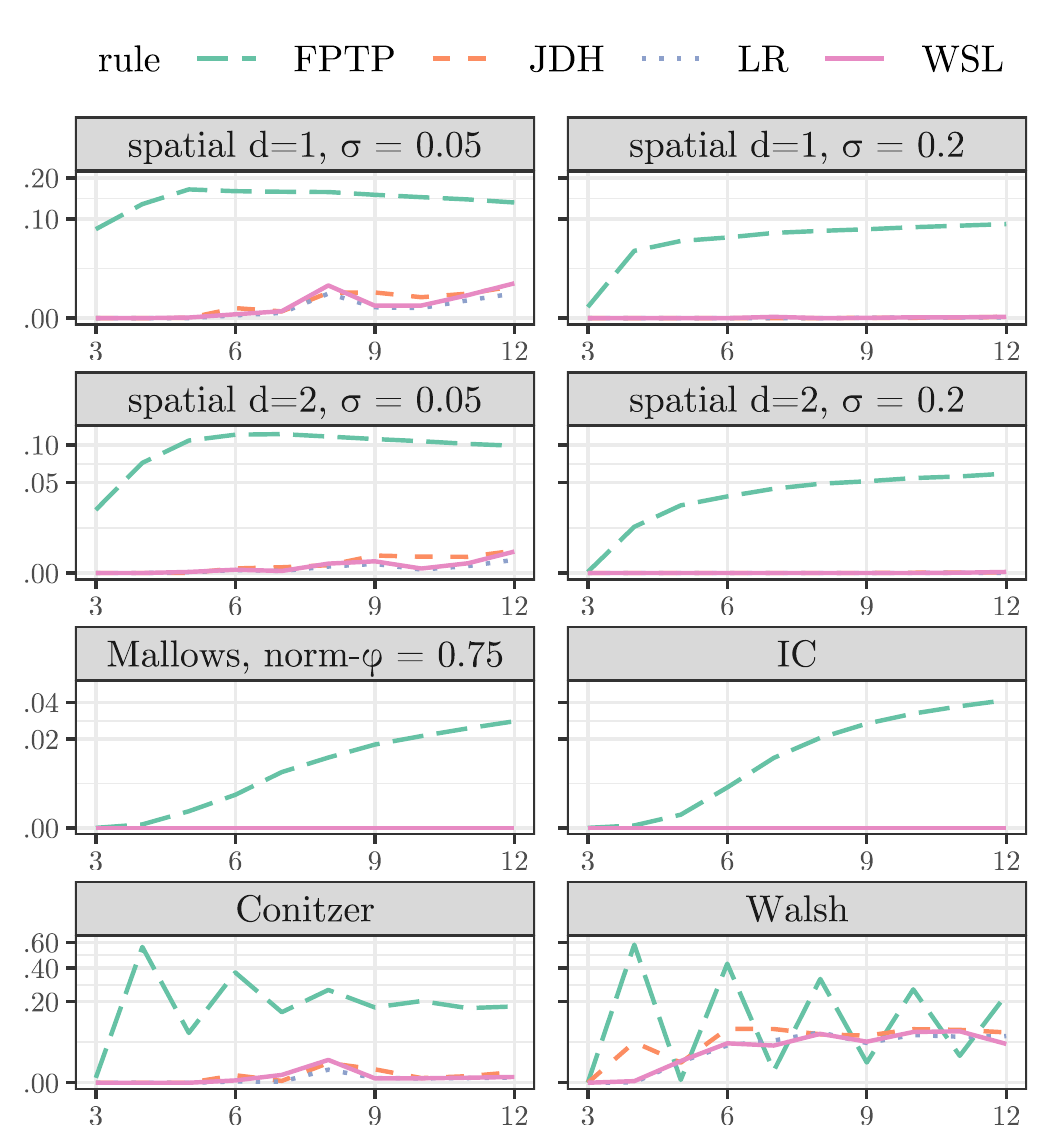}};
        \node[below=of img, node distance=0cm, yshift=1cm] {$p$ (number of parties)};
        \node[left=of img, node distance=0cm, rotate=90, anchor=center,yshift=-1.1cm] {$\Phi$ (spoiler susceptibility)};
    \end{tikzpicture}
    \caption{Spoiler susceptibility of electoral rules -- apportionment on the basis of Borda scores. Committee size, $k = 15$, is constant in each district. Note that the vertical scale differs between rows.}
    \label{fig:results15b}
\end{figure}

See Figures \ref{fig:results10} to \ref{fig:results15b} on the following page.

\clearpage

\FloatBarrier
\newpage



{
\bibliography{bib, spoilers}
}


\end{document}